\documentclass[12pt]{article}

\pdfoutput=1

\usepackage{graphicx}

\def\Title#1{\begin{center} {\Large #1 } \end{center}}
\def\Author#1{\begin{center}{ \sc #1} \end{center}}
\def\Address#1{\begin{center}{ \it #1} \end{center}}

\newenvironment{Abstract}{\begin{quotation}  }{\end{quotation}}
\newenvironment{Presented}{\begin{quotation} \begin{center} 
             PRESENTED AT\end{center}\bigskip 
      \begin{center}\begin{large}}{\end{large}\end{center} \end{quotation}}

\usepackage{defs}
\usepackage{booktabs}
\usepackage{placeins}
\usepackage[colorlinks=true,urlcolor=blue,linkcolor=blue,citecolor=blue]{hyperref}
\usepackage{caption}
\usepackage{subcaption}
\usepackage{cite}

\renewenvironment{thebibliography}[1]{
  \begin{oldthebibliography}{#1}
    \setlength{\itemsep}{0.3em}
    \setlength{\parskip}{0.3em}
}{
  \end{oldthebibliography}
}

\begin{document}

\begin{titlepage}
\vfill
\Title{Searches for rare top quark production and decay processes with the ATLAS experiment}
\vfill
\Author{William George~$^{a}$, on behalf of the ATLAS Collaboration\footnote{Copyright 2023 CERN for the benefit of the ATLAS Collaboration. CC-BY-4.0 license.}}
\Address{$^{a}$ University of Birmingham, West Midlands, England, UK}
\vfill
\begin{Abstract}
The large integrated luminosity collected by the ATLAS detector at the LHC provides the opportunity to probe the presence of new physics that could enhance the rate of very rare processes in the standard model (SM). The LHC can therefore gain considerable sensitivity for flavour changing neutral current (FCNC) and charged lepton flavour violating (CLFV) interactions of the top quark. In the SM, FCNCs are highly suppressed while CLFV interactions are forbidden
so any measurable branching ratio for such a process is an indication of new physics. 
The production of four top quarks is a rare SM process which also provides sensitivity to new physics processes. 
In this contribution, the latest results of searches for these processes are presented.
\end{Abstract}
\vfill
\begin{Presented}
DIS2023: XXX International Workshop on Deep-Inelastic Scattering and
Related Subjects, \\
Michigan State University, USA, 27-31 March 2023 \\
     \includegraphics[width=9cm]{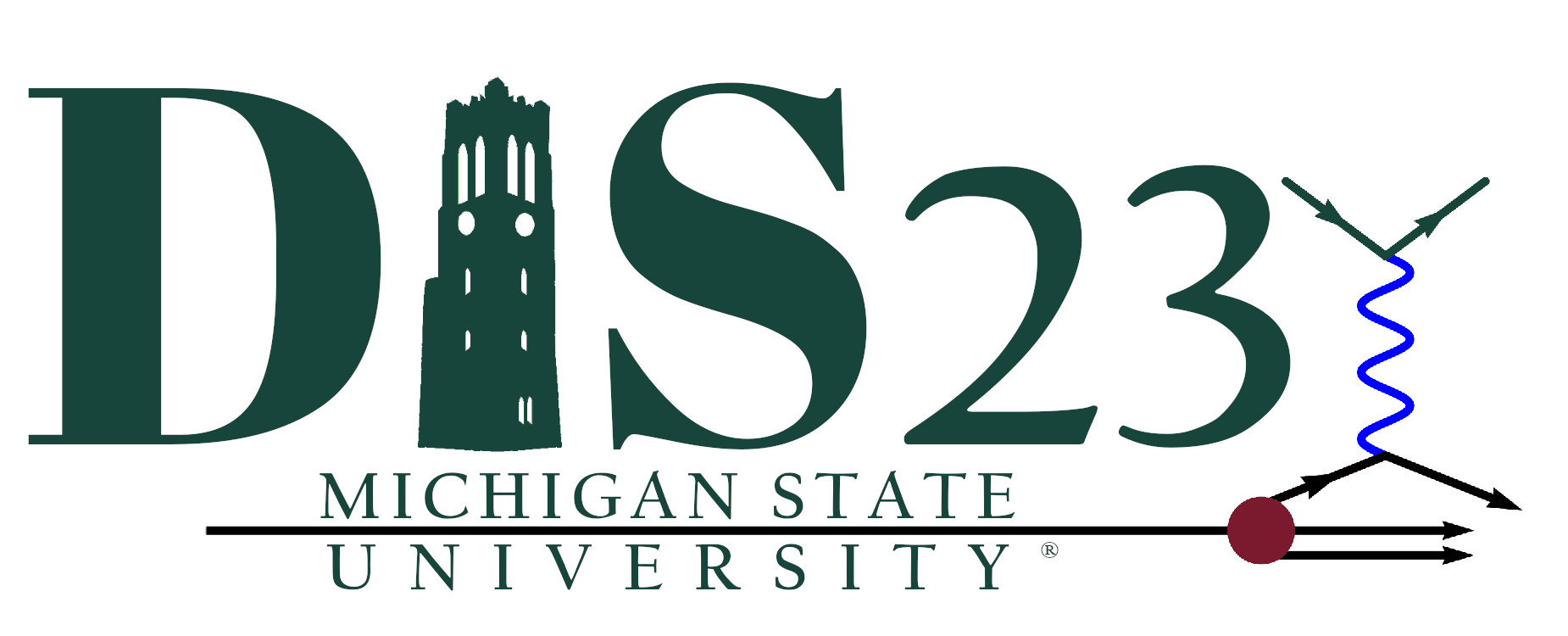}
\end{Presented}
\vfill
\end{titlepage}

\section{Introduction}

The top quark is the heaviest known elementary particle.
It couples strongly to the Higgs boson with a Yukawa coupling of order unity and may have large couplings to hypothetical new particles.
The ATLAS run 2 (2015-2018) data set provides a huge sample of top quark physics events which allow measurements of rare Standard Model (SM) processes as well as direct searches for new physics. This contribution presents the results of recent rare and beyond the SM (BSM) searches made by the ATLAS experiment~\cite{ATLAS-Experiment} at the Large Hadron Collider (LHC) using the full run 2 proton-proton collision data set of 139~\ifb{}.
The observation of four top quark production (\ft{}) is discussed, along with searches for flavour-changing neutral-current (FCNC) interactions of the top quark in $tqg$, $tq\gamma$, $tqZ$ and $tqH$ vertices ($q = \{u,c\}$), and charged lepton ﬂavour violating (CLFV) interactions via a \tmtq{} coupling.

\section{Observation of four top quark production}

The production of \ft{} is a very rare process with a SM cross-section of $\sigma^{SM}_{t\bar{t}t\bar{t}} = 12.0\pm2.4$ fb~\cite{Frederix-NLO-corrections}, yet many BSM models offer new production channels which could enhance the cross-section relative to the SM prediction.

ATLAS has performed a measurement of this process using final states with a single lepton or two oppositely-charged leptons ($1\ell/2\ell$OS)~\cite{four-tops-evidence}.
This channel features a large irreducible background from $t\bar{t}$ produced in association with heavy flavour jets.
Events are therefore categorised by jet multiplicity and $b$-tagging purity to resolve the different flavour components and kinematic reweighting is employed to correct for mis-modelling. 
A boosted decision tree (BDT) is used to discriminate between signal and background before a profile likelihood fit is used to extract the \ft{} cross-section.
A combination is performed with a previous measurement in same-charge dilepton and trilepton ($2\ell$SS$/3\ell$) final states~\cite{four-tops-ml-evidence} to yield a result of $\sigma_{t\bar{t}t\bar{t}} = 24\pm4(\mathrm{stat.})^{+5}_{-4}(\mathrm{syst.})$ fb, corresponding to an observed (expected) significance of 4.7$\sigma$ (2.6$\sigma$).

\begin{figure}
    \centering
    \begin{subfigure}[b]{0.39\textwidth}
         \centering
         \includegraphics[width=0.9\textwidth]{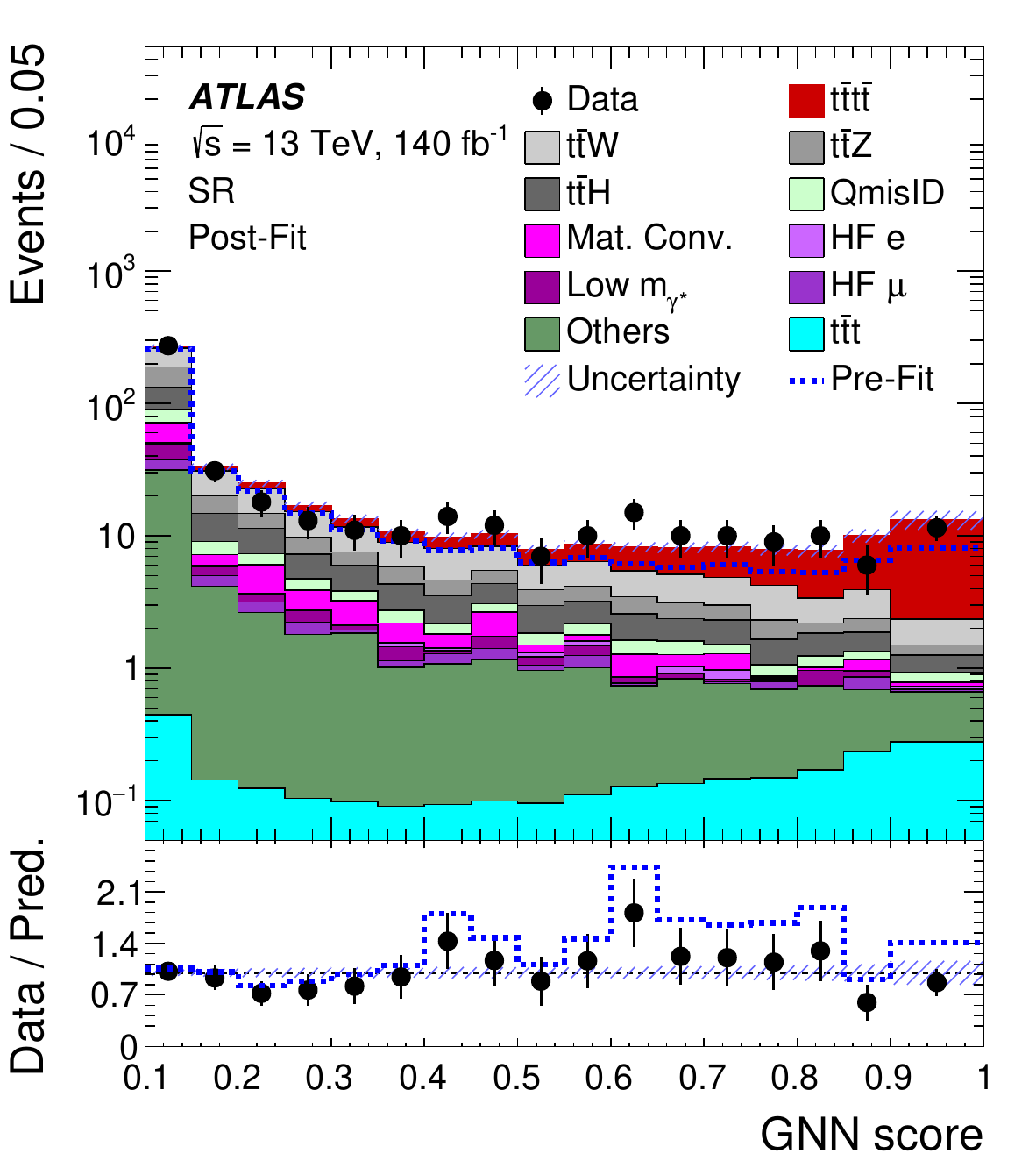}
         \caption{}
         \label{fig:four-tops_sr}
    \end{subfigure}
    \hfill
    \begin{subfigure}[b]{0.59\textwidth}
         \centering
         \includegraphics[width=\textwidth]{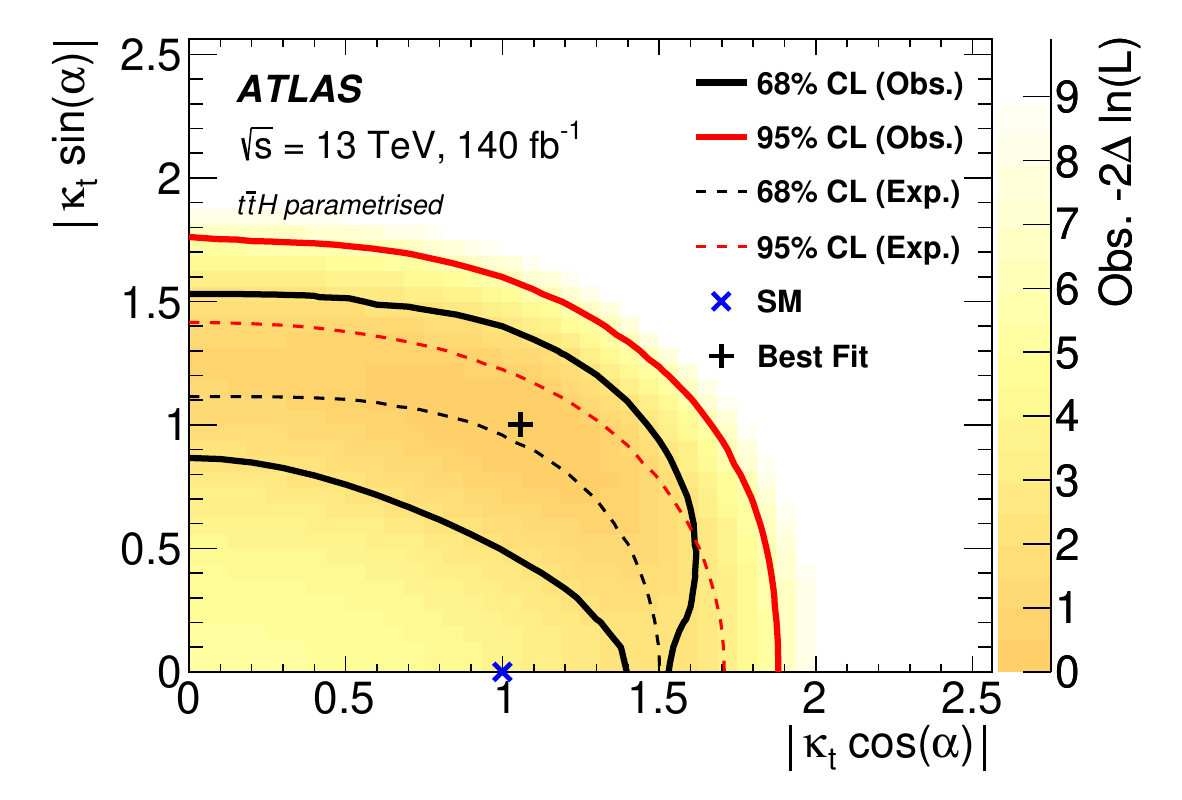}
         \caption{}
         \label{fig:four-tops_yt}
    \end{subfigure}
     \caption{Distribution of the NN discriminant in the signal region of the \ft{} analysis after the profile likelihood fit (left); constraints on the CP properties of the top Yukawa coupling (right)~\cite{four-tops-observation}.}
     \label{fig:four-tops}
\end{figure}

More recently, a refined measurement has been made in the $2\ell$SS$/3\ell$ channel~\cite{four-tops-observation}.
The analysis features data-driven background estimates of fake/non-prompt lepton backgrounds using dedicated control regions (CRs) and a jet multiplicity-dependent correction to the $t\bar{t}W$ background.
A graph neural network (NN) is used to discriminate signal from background and a profile likelihood fit is performed using the output score, shown in Figure~\ref{fig:four-tops_sr}.
The outcome of the measurement is $\sigma_{t\bar{t}t\bar{t}} = 22.7^{+4.7}_{-4.4}(\mathrm{stat.})^{+4.6}_{-3.4}(\mathrm{syst.})$ fb, corresponding to an observed (expected) significance of 6.1$\sigma$ (4.3$\sigma$).
The precision of the measurement is limited by the modelling of the signal and by the data-driven estimate of the $t\bar{t}W$ background.
The result lends itself to a number of interesting reinterpretations, including placing constraints on the CP properties of the top Yukawa coupling, shown in Figure~\ref{fig:four-tops_yt}.
A limit is also set on the rare three top quark production process and a number of four-fermion effective field theory (EFT) operators can be probed~\cite{four-tops-observation}.

\section{Search for FCNC interactions of the top quark}

FCNCs are forbidden at tree level in the SM and are heavily suppressed at loop level through the GIM mechanism. 
Nevertheless, a wide variety of BSM theories predict FCNCs with much larger rates, some of which may already be observable at the LHC~\cite{snowmass-2013-top-fcncs}.
Such results can also be interpreted in the context of EFT to constrain model-independent Wilson coefficients.
In all cases a binned proﬁle likelihood ﬁt is performed to extract the most likely signal and background normalisations.


A search for a FCNC \tqg{} vertex has been performed using a single lepton selection targeting events in which a single top quark is produced~\cite{fcnc-tqg}.
A data-driven estimate is used to model the background with fake leptons arising from multi-jet events.
To discriminate between signal events and background, NNs are employed, making use of the reconstructed top quark kinematics.
In the absence of an observed signal, exclusion limits are set on the $tug$ and $tcg$ signal processes at 95\% confidence level (CL).
The exclusion limits placed on $\mathcal{B}(t\to qg)$, and the branching ratios of the other top FCNC searches, are shown in Table~\ref{tab:fcnc_br_limits}.
The fit is systematically limited with dominant uncertainties in the $tug$ and $tcg$ channel from limited MC statistics and signal parton shower modelling, respectively.


A similar search has been performed for a FCNC \tqy{} coupling, but using both single top quark production and \ttbar{} decay~\cite{fcnc-tqy}.
Data-driven corrections are applied to the rate of electrons and hadrons mis-reconstructed as photons.
A multiclass NN is employed to classify events into two signal categories (production and decay) or background; the outputs of the NN are then combined into single discriminant to which a binned profile likelihood fit is performed.
Dedicated CRs for the main backgrounds with prompt photons from $t\bar{t}\gamma$ and $W\gamma$+jets are included in the fit.
No significant excess of events is observed in the data and the resulting exclusion limits can be seen in Table~\ref{tab:fcnc_br_limits}.
In both $tu\gamma$ and $tc\gamma$ analysis channels the dominant uncertainties come from limited statistics and the SM $tq\gamma$ cross-section.


In the \tqz{} search both top production and decay are included as signal and a low-background trilepton selection is used~\cite{fcnc-tqz}.
Events resembling the signal topology are split into two signal regions (SRs) targeting production and decay processes based on the reconstructed top quark mass.
BDTs are employed to resolve signal from background using the reconstructed kinematics.
A binned profile likelihood fit is performed on the SRs and additional CRs to control the dominant diboson and $t\bar{t}Z$ backgrounds, as well as \ttbar{} events with fake leptons.
The data are seen to be consistent with the background model and the corresponding exclusion limits on the process are shown in Table~\ref{tab:fcnc_br_limits}.
The uncertainty is dominated by limited statistics. 

\begin{figure}[htbp]
    \centering
    \begin{subfigure}[b]{0.48\textwidth}
         \centering
         \includegraphics[width=0.65\textwidth]{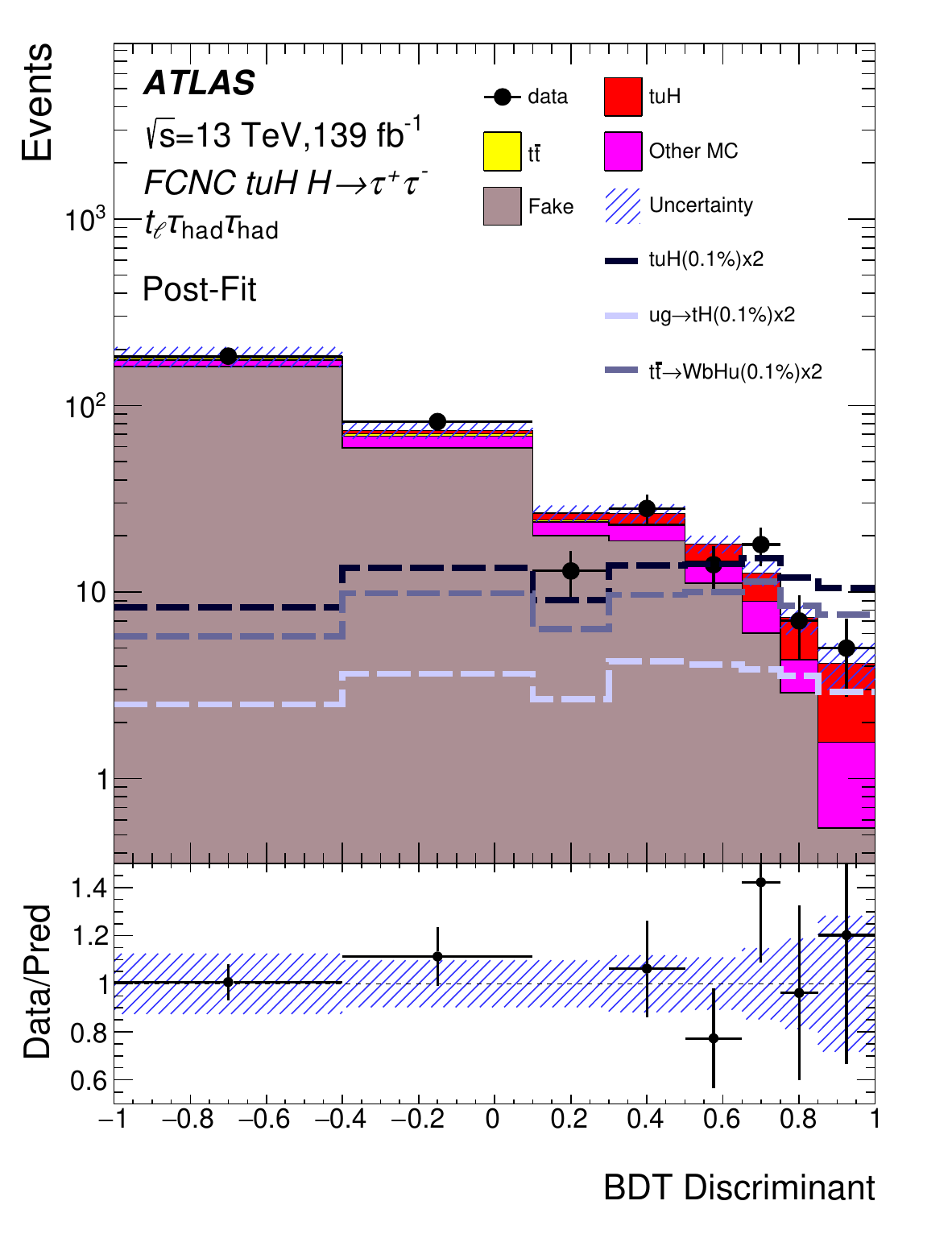}
         \caption{}
         \label{fig:tqH_SR_u}
    \end{subfigure}
    \hfill
    \begin{subfigure}[b]{0.48\textwidth}
         \centering
         \includegraphics[width=0.65\textwidth]{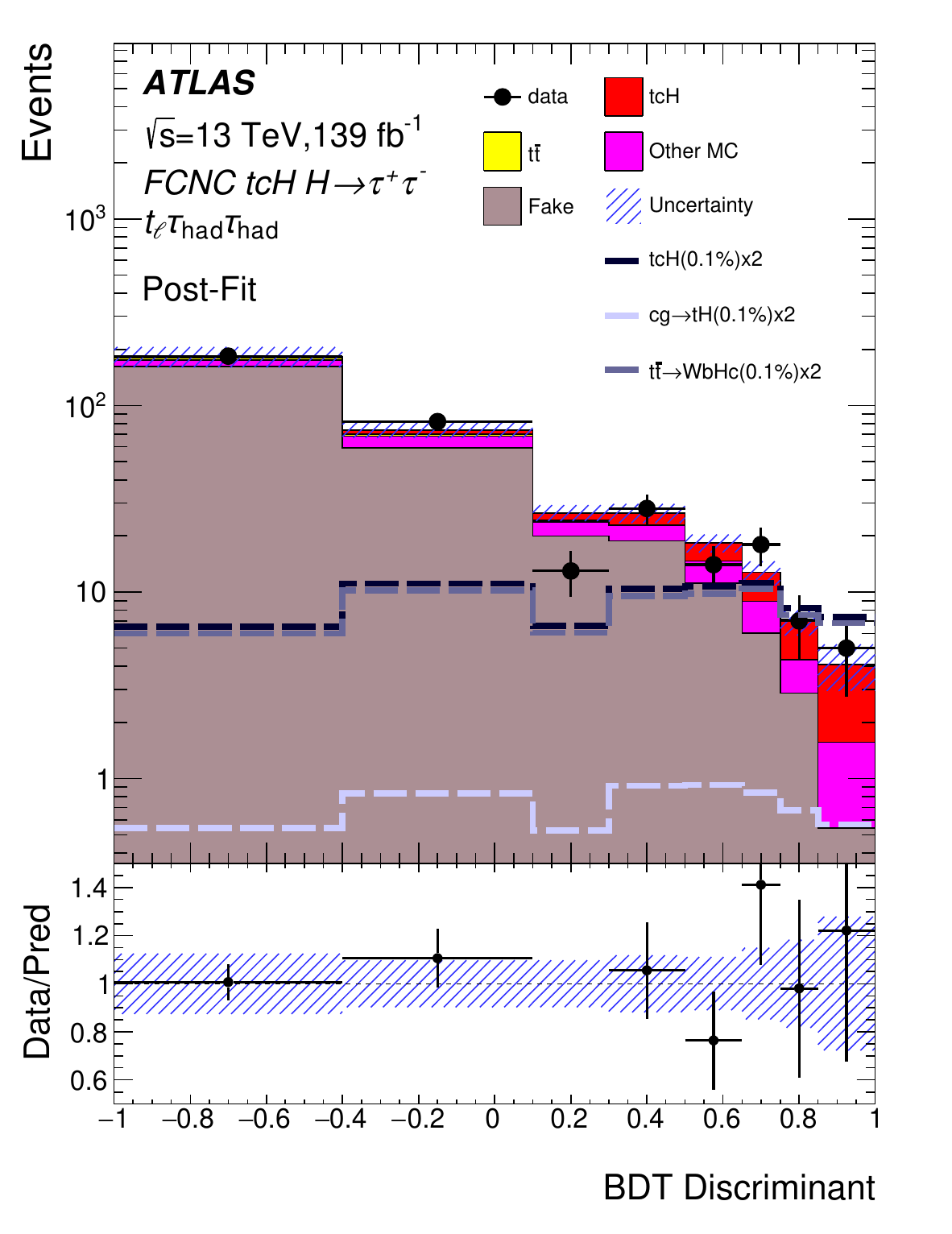}
         \caption{}
         \label{fig:tqH_SR_c}
    \end{subfigure}
     \caption{BDT output distributions for most sensitive channel of \tqh{} search after the profile likelihood fit for $tuH$ (left) and $tcH$ (right) searches~\cite{fcnc-tqh}.}
     \label{fig:tqH_SR}
\end{figure}


The search for a \tqh{} coupling also considers both single top production and \ttbar{} decay, requiring the Higgs boson to decay as $H\to\tau\tau$~\cite{fcnc-tqh}.
Many SRs are defined targeting different top and $\tau$ lepton decay channels.
Data-driven estimates of the background from fake hadronically-decaying taus (\tauhadvis{}) and fake/non-prompt light leptons are made.
BDTs are used for signal-background discrimination relying on event kinematics, including reconstruction of the top and $\tau$ decays where possible.
A profile likelihood fit is performed to 7 SRs and 6 CRs and a slight excess of 2.3$\sigma$ is observed compared to the expected background.
The excess is driven by the high BDT score region in the SR with the best signal sensitivity, shown in Figure~\ref{fig:tqH_SR}. 
The background modelling in validation regions is seen to be good while kinematic distributions are compatible with both the background shape or a small signal within the large statistical uncertainties which dominate the measurement.
The resulting exclusion limits on the branching ratio of the FCNC top decay are shown in Table~\ref{tab:fcnc_br_limits}.

\begin{table}[htbp]
\centering
\caption{Observed 95\% CL upper limits on branching ratios of FCNC top decays~\cite{fcnc-tqg,fcnc-tqy,fcnc-tqz,fcnc-tqh}. Results are shown separately for $\mathcal{B}(t\to uX)$ and $\mathcal{B}(t\to cX)$. For \tqy{} and \tqz{} the limits on left-handed couplings are shown.}
\small
\begin{tabular}{c|ccc}
\toprule
 & $\mathcal{B}(t\to uX)$ & $\mathcal{B}(t\to cX)$ & \\
\midrule
$\mathcal{B}(t\to qg)$ & $0.61\times10^{-4}$ & $3.7\times10^{-4}$ &  \\
$\mathcal{B}(t\to q\gamma)$ & $0.85\times10^{-5}$ & $4.2\times10^{-5}$ &  \\
$\mathcal{B}(t\to qZ)$ & $6.2\times10^{-5}$ & $13\times10^{-5}$ &  \\
$\mathcal{B}(t\to qH)$ & $6.9\times10^{-4}$ & $9.4\times10^{-4}$ &  \\
\bottomrule
\end{tabular}
\label{tab:fcnc_br_limits}
\end{table}

\FloatBarrier
\section{Search for CLFV interactions of the top quark}

Lepton flavour is conserved in the SM due to an accidental symmetry stemming from the absence of a right-chiral neutrino field. 
However, lepton flavour violation features in several BSM models, such as Grand Unified theories in which leptoquarks can arise as the gauge bosons of the new symmetry.

A search for a CLFV \tmtq{} interaction has been performed by ATLAS, which represents the first direct search for such a coupling~\cite{clfv}.
The signal process under study includes both production and decay processes, $qg \to t\ell\ell^{'}$ and $t\bar{t}\to (\ell\ell^{'}q)((W\to \ell\nu) b)$, respectively, where $\ell$ is a muon or tau lepton.
The signal is characterised by two isolated muons, one \tauhadvis{}, and at least one jet, of which exactly one is required to be $b$-tagged.
Any possible signal in the data is extracted by a binned profile likelihood fit, which also corrects the rate of the non-prompt muon background using a floating normalisation factor.
Scale factors derived in a dedicated CR are used to correct the modelling of the fake \tauhadvis{} background before the fit.
No significant excess is observed above the expected background and stringent limits are placed at 95\% CL on the branching ratio $\mathcal{B}(t\to\mu\tau q) < 10.8\times10^{-7}$ and on Wilson coefficients corresponding to the four-fermion operators parametrising the CLFV processes, shown in Table~\ref{tab:clfv_eft_wc_limits}.
The result is found to be statistically limited.

\begin{table}[htbp]
\centering
\caption{Observed 95\% CL upper exclusion limits on Wilson coefficients corresponding to four-fermion EFT operators which could introduce a CLFV \tmtq{} interaction~\cite{clfv}, and existing limits from Ref.~\cite{Chala-four-fermion-top} (previous).}
\footnotesize
\renewcommand{\arraystretch}{1.2}
\begin{tabular}{c|cccccccc}
\toprule
 & \multicolumn{8}{c}{\textbf{95\% CL upper limits on Wilson coefficients \quad $c / \Lambda^{2}$ [$\textrm{TeV}^{-2}$] }} \\[0.5ex]
 & $c^{-(ijk3)}_{lq}$ & $c^{(ijk3)}_{eq}$ & $c^{(ijk3)}_{lu}$ & $c^{(ijk3)}_{eu}$ & $c^{1(ijk3)}_{lequ}$ & $c^{1(ij3k)}_{lequ}$ & $c^{3(ijk3)}_{lequ}$ & $c^{3(ij3k)}_{lequ}$ \\[1.0ex] 
 \midrule
\textbf{Previous (u)}~\cite{Chala-four-fermion-top} & 12 & 12 & 12 & 12 & 26 & 26 & 3.4 & 3.4 \\
\textbf{Observed (u)} & 0.49 & 0.47 & 0.46 & 0.48 & 0.51 & 0.51 & 0.11 & 0.11 \\
\midrule
\textbf{Previous (c)}~\cite{Chala-four-fermion-top} & 14 & 14 & 14 & 14 & 29 & 29 & 3.7 & 3.7 \\
\textbf{Observed (c)} & 1.7 & 1.6 & 1.6 & 1.6 & 1.9 & 1.9 & 0.37 & 0.37 \\
\bottomrule
\end{tabular}
\label{tab:clfv_eft_wc_limits}
\end{table}

\FloatBarrier
\section{Conclusion}

The ATLAS experiment at the LHC has performed a number of interesting searches for rare and BSM top quark interactions making use of the full run 2 data set, including a first observation of \ft{} production.
Many of these results, particularly those using multi-lepton selections, are statistically limited so further improvements can be expected from run 3 and the HL-LHC era.



\begin{thebibliography}{15}

\bibitem{ATLAS-Experiment}
ATLAS Collaboration, \href{https://iopscience.iop.org/article/10.1088/1748-0221/3/08/S08003}{JINST 3 (2008) S08003}.

\bibitem{Frederix-NLO-corrections}
Frederix, R. \textit{et al.}, \href{https://doi.org/10.1007/JHEP02(2018)031}{JHEP 31 (2018) 496}.

\bibitem{four-tops-evidence}
ATLAS Collaboration, \href{https://link.springer.com/article/10.1007/JHEP11(2021)118}{JHEP 11 (2021) 118}.

\bibitem{four-tops-ml-evidence}
ATLAS Collaboration, \href{https://link.springer.com/article/10.1140/epjc/s10052-020-08509-3}{Eur. Phys. J. C 80 (2020) 1085}.

\bibitem{four-tops-observation}
ATLAS Collaboration, \href{https://link.springer.com/article/10.1140/epjc/s10052-023-11573-0}{Eur. Phys. J. C 83 (2023) 496}.

\bibitem{snowmass-2013-top-fcncs}
Agashe, K. \textit{et al.}, \href{https://arxiv.org/abs/1311.2028}{arXiv:1311.2028 [hep-ph] (2013)}.

\bibitem{fcnc-tqg}
ATLAS Collaboration, \href{https://link.springer.com/article/10.1140/epjc/s10052-022-10182-7}{Eur. Phys. J. C 82 (2021) 334}.

\bibitem{fcnc-tqy}
ATLAS Collaboration, \href{https://www.sciencedirect.com/science/article/pii/S0370269322005135}{Physics Letters B 842 (2022) 137379}.

\bibitem{fcnc-tqz}
ATLAS Collaboration, \href{https://arxiv.org/abs/2301.11605}{arXiv:2301.11605 [hep-ex] (2023)}, submitted to Phys. Rev. D.

\bibitem{fcnc-tqh}
ATLAS Collaboration, \href{https://arxiv.org/abs/2208.11415}{arXiv:2208.11415 [hep-ex] (2022)}, submitted to JHEP.

\bibitem{clfv}
ATLAS Collaboration, \href{http://cds.cern.ch/record/2845451}{ATLAS-CONF-2023-001, http://cds.cern.ch/record/2845451}.

\bibitem{Chala-four-fermion-top}
Chala, M. \textit{et al.}, \href{https://link.springer.com/article/10.1007/JHEP04(2019)014}{JHEP 4 (2019) 14}.

\end{thebibliography}
\end{document}